\documentclass[twocolumn]{emulateapj}

\usepackage{amsmath}
\usepackage{url}
\usepackage{color}
\usepackage{soul}
\usepackage{ulem}

\shortauthors{
Wong et al.
}

\shorttitle{
Hard X-ray Emission of M87
}
\slugcomment{Astrophysical Journal Letters}

\begin{document}
\title{
Hard X-ray Emission from the M87 AGN Detected with {\it NuSTAR}
}

\author{
Ka-Wah Wong\altaffilmark{1,2},
Rodrigo S. Nemmen\altaffilmark{3},
Jimmy A. Irwin\altaffilmark{4},
and
Dacheng Lin\altaffilmark{5},
}

\altaffiltext{1}{Eureka Scientific, Inc., 2452 Delmer Street Suite 100,
Oakland, CA 94602-3017, USA
}
\altaffiltext{2}{Department of Physics and Astronomy, Minnesota State
University, Mankato, MN 56001, USA
}
\altaffiltext{3}{Universidade de S\~ao Paulo, Instituto de Astronomia, 
Geof\'{\i}sica e Ci\^encias Atmosf\'ericas, S\~ao Paulo, SP 05508-090, 
Brazil
}
\altaffiltext{4}{Department of Physics and Astronomy, University of
Alabama, Box 870324, Tuscaloosa, AL 35487, USA
}
\altaffiltext{5}{Space Science Center, University of New Hampshire,
Durham, NH 03824, USA
}
\email{kw6k@email.virginia.edu}


\begin{abstract}

M87 hosts a 3--6 billion solar mass black hole with a remarkable 
relativistic jet that has been regularly monitored in radio to TeV 
bands. However, hard X-ray emission $\gtrsim$10\,keV, which would be 
expected to primarily come from the jet or the accretion flow, had never 
been detected from its unresolved X-ray core.  We report {\it NuSTAR} 
detection up to 40\,keV from the the central regions of M87. Together 
with simultaneous {\it Chandra} observations, we have constrained the 
dominant hard X-ray emission to be from its unresolved X-ray core, 
presumably in its quiescent state.  The core spectrum is well fitted by 
a power law with photon index $\Gamma=2.11^{+0.15}_{-0.11}$.  The 
measured flux density at 40\,keV is consistent with a jet origin, 
although emission from the advection-dominated accretion flow cannot be 
completely ruled out.  The detected hard X-ray emission is significantly 
lower than that predicted by synchrotron self-Compton models introduced 
to explain emission above a GeV.

\end{abstract}

\keywords{
accretion, accretion disks ---
black hole physics ---
galaxies: elliptical and lenticular, cD ---
galaxies: individual (M87) ---
galaxies: nuclei ---
X-rays: galaxies
}

\section{Introduction}
\label{ion_sec:intro}

One of the best-studied active galactic nuclei (AGNs) is located at the 
center of the nearby radio galaxy M87.  The giant elliptical galaxy 
hosts a $\sim$3--6 billion solar mass supermassive black hole 
\citep[SMBH;][]{MMA+97,GT09}, and contains the first extragalactic 
astrophysical jet discovered \citep{Cur18}.  Its proximity 
\citep[$D$=16\,Mpc, 1\arcsec=78\,pc;][]{TDB+01} allows us to resolve 
the arcsecond-scale relativistic jet near the nucleus region with radio 
\citep[$\sim$40 micro-arcsecond: 230\,GHz high frequency 
VLBI;][]{Doe+12}, optical (0.1\,arcsec: {\it HST}), and X-ray 
(sub-arcsecond: {\it Chandra}) observations.  It is also a 
very-high-energy (VHE; $>$100\,GeV) source detected with the {\it Fermi}, 
H.E.S.S., VERITAS, and MAGIC $\gamma$-ray telescopes \citep[see,][and 
references therein]{Abr+12}, and one of the few radio galaxies with TeV 
$\gamma$-rays detected but not strongly beamed.\footnote{The others are 
Cen A, NGC~1275, and IC~310.}  This provides a different scientific 
angle to study high-energy particle acceleration mechanisms in AGNs with 
misaligned jets.

Multi-wavelength observations resolve the M87 central region into 
different components at different angular scales: a core where the SMBH 
is located, a jet, and multiple knots along the jet direction 
\citep[e.g.,][]{BSM99,JBL99,KLH+07,HCS+09}.  The most obvious features 
on {\it Chandra} X-ray images are the core and the bright knots 
(hereafter, core means the central source unresolved by {\it Chandra} at 
the sub-arcsecond scale, following the definition in Figure~1 of 
\citet{HCB+06}; see also the right panel of Figure~1 below).  Outbursts 
at different wavelengths have been seen from both the core and knots 
\citep{HCS+09,Abr+12}.  The locations of the highest energy burst 
(GeV--TeV) are not very clear.  For example, it had been suggested that 
the HST-1 knot outburst is responsible for the TeV outburst seen in 2005 
\citep{Aha+06}.  However, later observations, particularly 
with {\it Chandra}, support that the TeV emission comes from the 
unresolved {\it Chandra} core (Abramowski et al.\ 2012).  The exact 
origin of the different energy emission is still under debate.  Studying 
the full spectral energy distribution (SED) at different states 
(quiescence or flare) of the core and knots will hopefully help us to 
distinguish different models \citep[e.g.,][]{NSE14,dBS+15,PFM+16}.

X-rays penetrate the dust-free M87 nucleus \citep[e.g.,][]{PSR+01} 
directly from the inner most accretion regions of the flow, allowing us 
to study physics down to the event horizon scale.  Because it is a 
low-luminosity AGN (LLAGN) with an Eddington ratio of $3.6\times 
10^{-6}$ \citep{PFM+16}, the SMBH is believed to be accreting as an 
advection-dominated accretion flow \citep[ADAF; see, e.g.,][for a 
review]{YN14}.  The X-ray emission from the M87 core is believed to come 
from either the ADAF or the (unresolved) jet, therefore studying the 
X-ray emission will provide constraints on these models \citep{NSE14}.  
Although X-ray emission $\lesssim$10\,keV has been studied in detail, 
hard X-rays above that threshold had never been detected from the 
core.\footnote{\citet{dBS+15} detected hard X-ray emission observed with 
{\it Suzaku} in 2006, which likely comes from HST-1.}  In this Letter, 
we report the detection of hard X-ray emission from the core up to 
40\,keV, with simultaneous {\it NuSTAR} and {\it Chandra} observations.  
Errors are given at the 90\% confidence level unless otherwise 
specified.

\begin{figure*}
\includegraphics[width=0.33\textwidth]{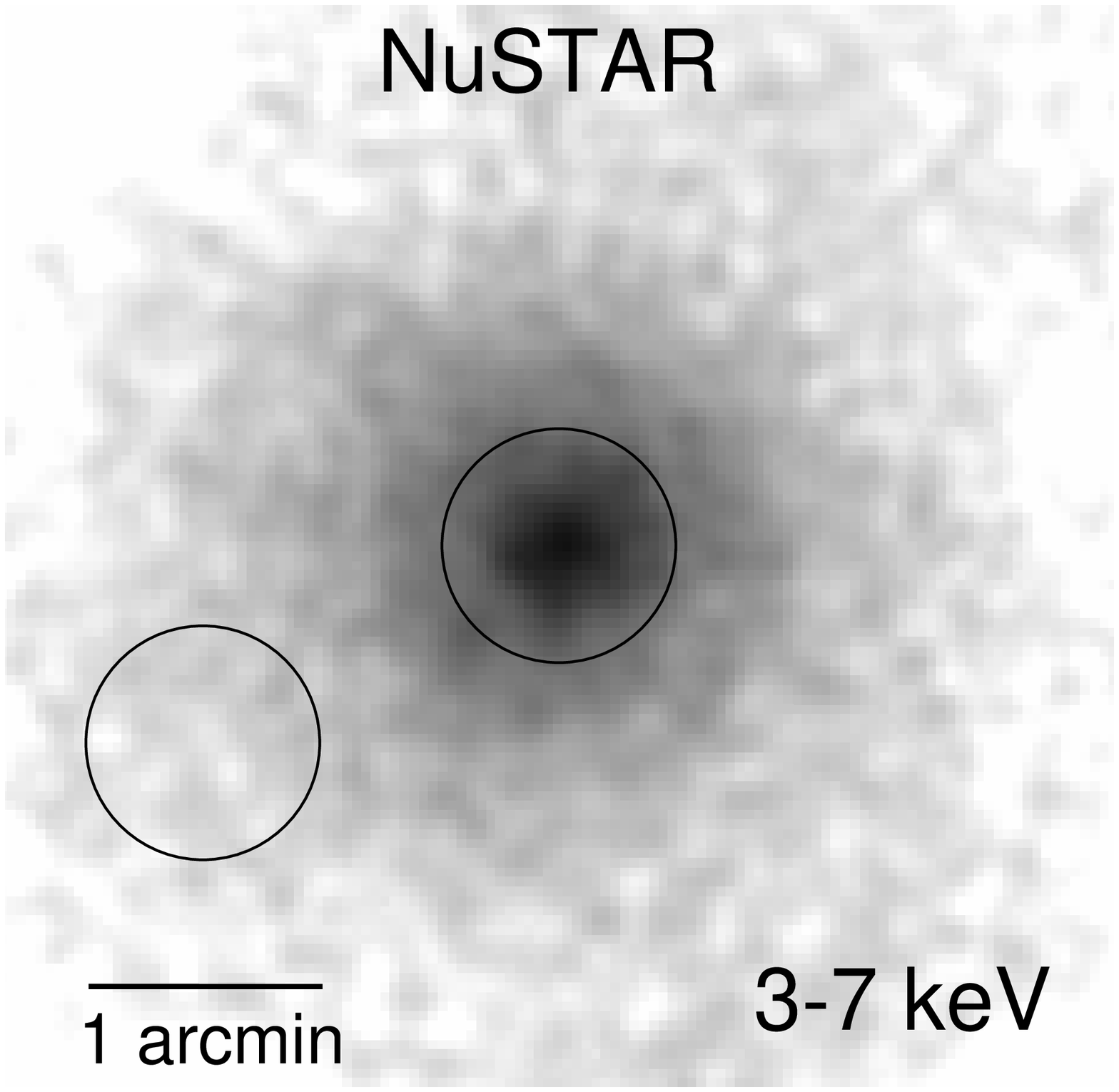}
\includegraphics[width=0.33\textwidth]{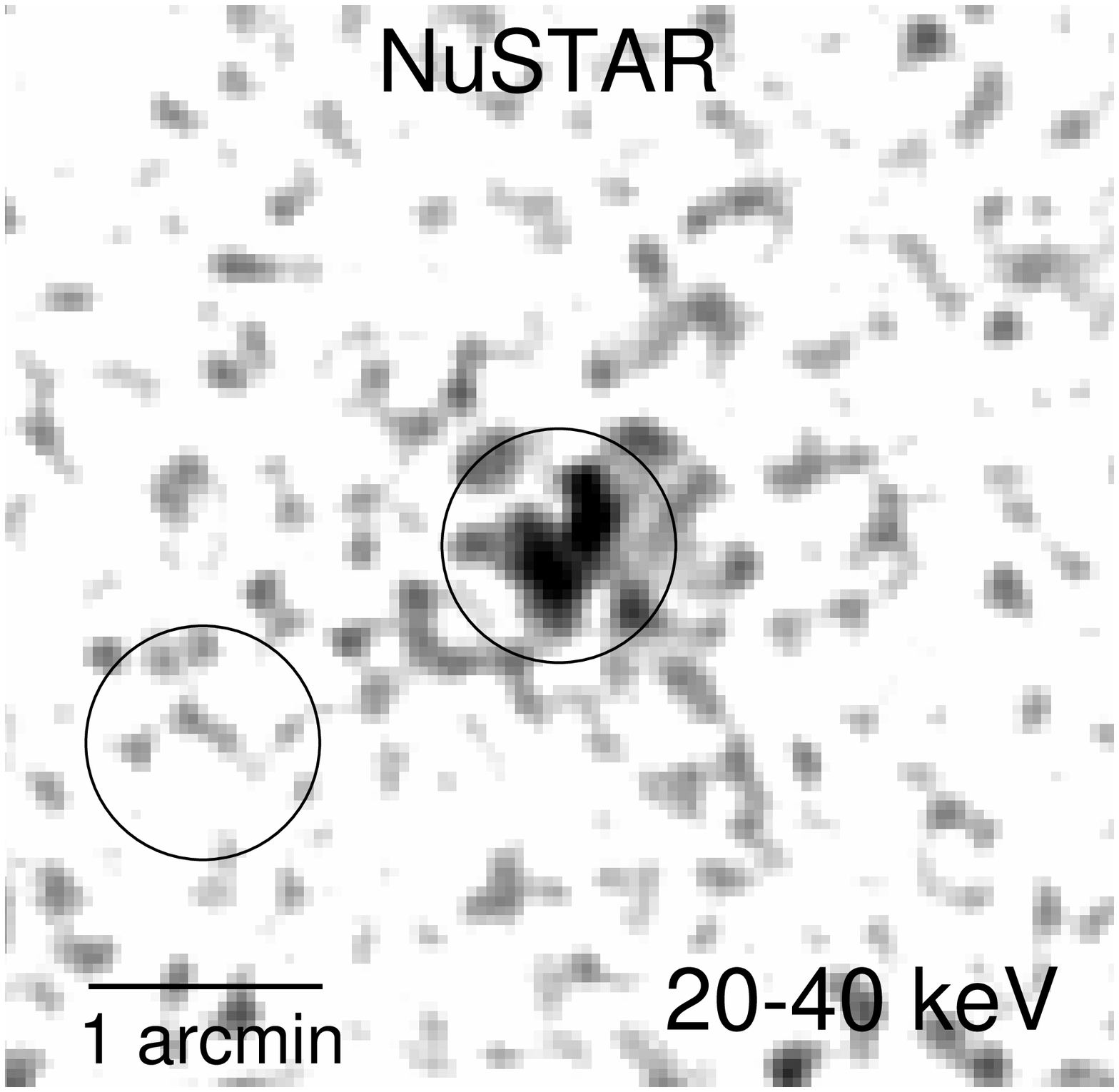}
\includegraphics[width=0.33\textwidth]{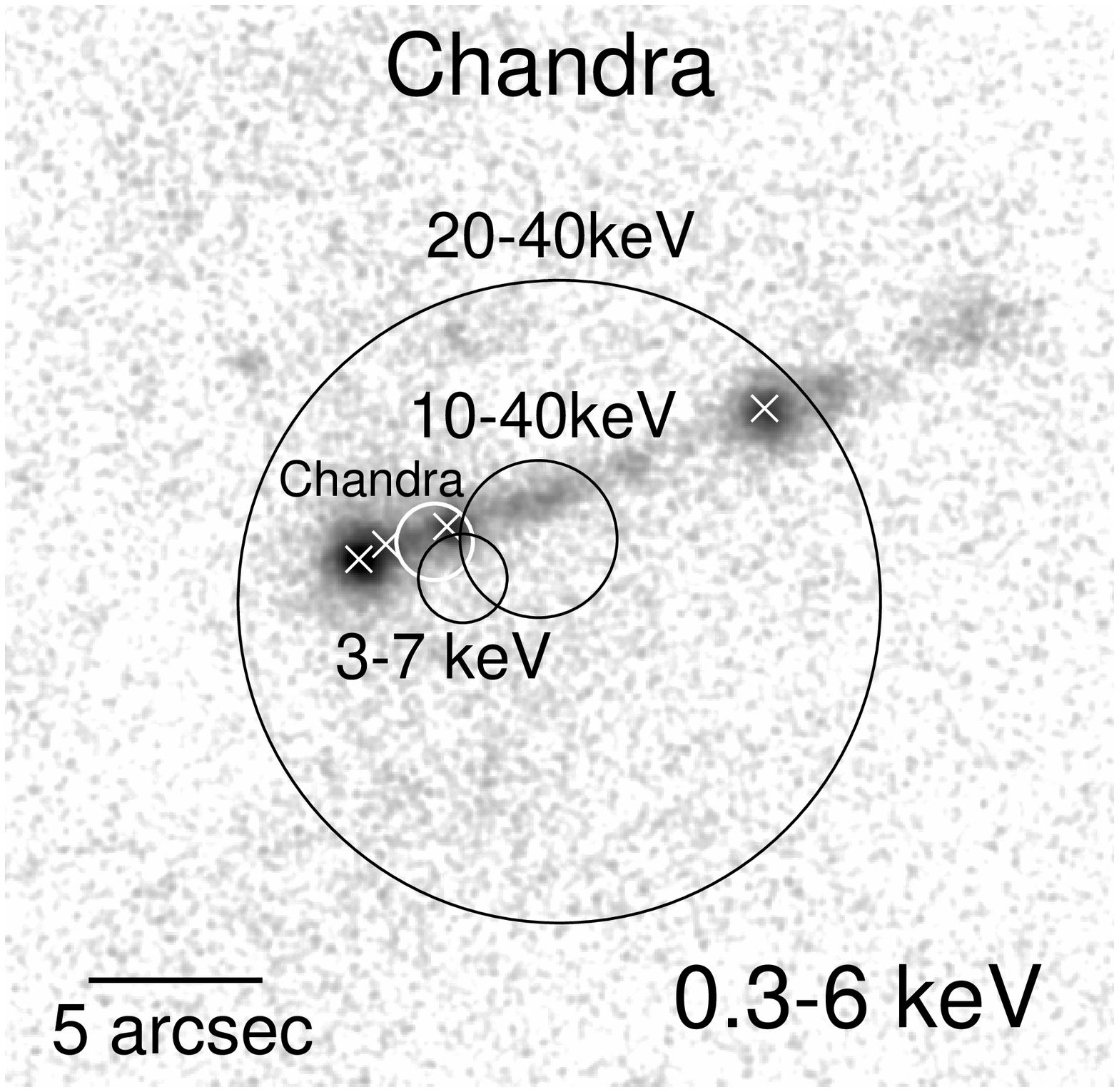}
\caption{Smoothed 3--7\,keV (left) and 20--40\,keV (middle) {\it NuSTAR} 
images of M87.  The circle at the center of the image indicates the 
30\arcsec\ radius spectral extraction region.  The circle in the lower left is
the background spectral extraction region with the same radius of 30\arcsec.
A smoothed 0.3--6\,keV {\it Chandra} image is shown in the right panel.
The four white crosses from the left to the right are the locations of the 
core, the HST-1 knot, knot D, and knot A, respectively.
The three black solid circles indicate the 3$\sigma$ position errors of 
the emission
peaks measured with {\it NuSTAR} in different energy bands.  
The white circle indicates the {\it Chandra} position determined 
using a smoothed image in the 3--7\,keV band (see the text).
}
\label{fig:image}
\end{figure*}

\section{X-ray Observations}
\label{sec:obs}

M87 was observed with {\it NuSTAR} on 2017 February 15, April 11, and 
April 14 for 50, 24, and 22\,ks, respectively (ObsIDs: 60201016002, 
90202052002, and 90202052004). All data were reduced using the {\tt 
HEASoft} v6.21 and {\tt CALDB} version 20170616.  We have reprocessed 
the data using the {\tt nupipeline} script of the {\it NuSTAR} Data 
Analysis Software ({\tt NuSTARDAS}) package with the standard screening 
criteria. Four {\it Chandra} snapshots were taken during the three {\it 
NuSTAR} observations on 2017 February 15, February 16, April 11, April 
14 for 5, 5, 13, and 13\,ks, respectively (ObsIDs: 19457, 19458, 20034, 
and 20035). All the data were reprocessed using the {\it Chandra} 
Interactive Analysis of Observations ({\tt CIAO}) software version 4.9 
and the {\it Chandra} {\tt CALDB} version 4.7.4.

Figure~\ref{fig:image} shows the {\it NuSTAR} images in 3--7\,keV and 
20--40\,keV.  Each image was created by combining the three observations 
and the two detectors.  Relative astrometry was corrected by matching 
the centroids of the 3--7\,keV emission.  Hard X-ray emission in 
20--40\,keV is clearly detected at 7$\sigma$.

To quantify the spatial structure of the hard X-ray (20--40\,keV) 
emission, we fit the hard X-ray image with two 2D Gaussian models to 
represent the PSF core and larger PSF wing and a constant background 
model.  The fitting region is limited to a circular region of radius 
$\sim$2\arcmin\ centered near the peak of the emission.  The fitted FWHM 
is consistent with the 18\arcsec\ FWHM of the {\it NuSTAR} PSF, 
indicating that the hard X-ray emission is unresolved.  The location of 
the hard X-ray peak can be determined to about 9\arcsec\ at 3$\sigma$
for two parameters of interest, 
and the confidence region is shown in the right panel of 
Figure~\ref{fig:image}.  Using a wider hard band in 10--40\,keV, the 
error circle has a smaller radius of 2\farcs3. In contrast, the soft 
X-ray emission in 3--7\,keV is clearly extended due to the strong 
intracluster medium (ICM) emission.  We determined the location of the 
soft X-ray peak by fitting a 2D Gaussian plus a 2D $\beta$-model with a 
constant background.
We tied the centers of the first two azimuthally symmetric models and 
thawed all the remaining parameters.
The much smaller ($\sim$1\arcsec) confidence 
region is also shown in the figure.

The peak location of the 10--40\,keV hard X-ray emission is marginally 
consistent with the softer 3--7\,keV peak. Thus, the origin of the hard 
X-ray emission above 10\,keV is consistent with the origin of the softer 
emission (i.e., core/knots/jet emission contaminated by ICM). Deeper 
{\it NuSTAR} observations can potentially distinguish any structure to 
better than a few arcsec.

The location of the {\it NuSTAR} soft peak can be compared with that 
measured with {\it Chandra}.  We have smoothed a {\it Chandra} 3--7\,keV 
image with a 2D Gaussian to simulate the FWHM of the {\it NuSTAR} PSF 
(18\arcsec).  The confidence region of the smoothed {\it Chandra} image 
is located near knot D.  It is only slightly offset by about 1\arcsec\ 
from the {\it NuSTAR} soft peak, which is much smaller than the 
8\arcsec\ absolute astrometric uncertainty for {\it NuSTAR}.

We extracted {\it NuSTAR} spectra from the nuclear region of M87 
for all 
the three observations and created the corresponding response files for the 
point source.
The extraction region is 
circular with a 30\arcsec\ radius, which is close to the 29\arcsec\ 
half-power radius, centered at the {\it NuSTAR} 3--7\,keV peak 
(Figure~\ref{fig:image}). The background circular region with the same 
radius was chosen to be far enough from the center and also located on 
the same detector chip as the source.  The latter criterion is important 
to minimize instrumental background variations from chip to chip 
\citep{Wik+14a}. Since the individual spectra are consistent with each 
other, we combined all the {\it NuSTAR} spectral and response files 
using the {\tt FTOOL addspec}.

The unresolved {\it NuSTAR} X-ray emission is mainly contributed by the 
X-ray core, the jet and knots along the jet, the diffuse ICM, and 
unresolved low-mass X-ray binaries (LMXBs). With simultaneous 
observations and archival {\it Chandra} data, these individual 
components can be well constrained.  We first detected point sources 
using the {\it Chandra} data following the procedure described in 
\citet{WIS+14}.  To detect fainter sources, we also included two deeper 
{\it Chandra} observations taken in 2016 (ObsIDs: 18838 and 18856 for 56 
and 25\,ks, respectively).  We extracted {\it Chandra} spectra from the 
six observations with detected point sources inside the {\it NuSTAR} 
spectral extraction region that are unrelated to the core, jet, or 
knots.  Each spectral region is circular with a radius of 1\arcsec.  A 
corresponding local annular background with inner and outer radii of 
1\arcsec\ and 2\arcsec, respectively, is centered on each source.  The 
combined spectrum is used to constrain the LMXB component.

To constrain the jet component, we extracted {\it Chandra} spectra using 
a $19\farcs5\times3\arcsec$ rectangular region enclosing all the jet and 
knots (but not the unresolved {\it Chandra} core or HST-1).  A local 
background with two similar rectangular regions adjacent to the source 
region is used.

For both the core and the HST-1 knot, we extracted {\it Chandra} 
spectra with circular regions of 0\farcs4 radius centered on 
them 
and created the corresponding response files for the point sources.
For the core, a pie-shaped local background with inner and outer 
radii of 2\arcsec\ and 4\arcsec, respectively, centered on the core and 
away from HST-1 is used.  For HST-1, a circular region of 0\farcs4 
radius opposite from the core is chosen as its local background.  For 
the core, jet, and HST-1 spectra, only the simultaneous observations 
taken in 2017 were used.  The core and HST-1 were in their low states 
during the 2017 observations and their total pile-up fraction is not 
significant (3--7\%).\footnote{The pile-up effect on spectral slope,
determined by fitting a {\tt pileup} model, is always $\lesssim$ the 
statistical uncertainty.}

The remaining component is mostly the steady ICM emission.  Since the 
{\it NuSTAR} background spectrum also contains ICM emission (and a 
negligible contribution from LMXBs), we extracted {\it Chandra} ICM 
spectra from the same {\it NuSTAR} source and background regions to take 
into account this component.  The abovementioned {\it Chandra} 
observations were taken in the 1/8-subarray mode that do not cover the 
{\it NuSTAR} background region.  Therefore, we selected earlier {\it 
Chandra} archival data (ObsIDs: 352, 2707, 3717, and 8517) that cover 
both the {\it NuSTAR} source and background regions.  The total cleaned 
exposure time is $\sim$110\,ks, similar to that of the six {\it Chandra} 
observations used to extract the LMXB spectra.  We removed all the point 
sources detected with the same method used for the {\it Chandra} ICM source 
and background spectra.

\begin{figure}
\includegraphics[width=2.5truein, angle=270]{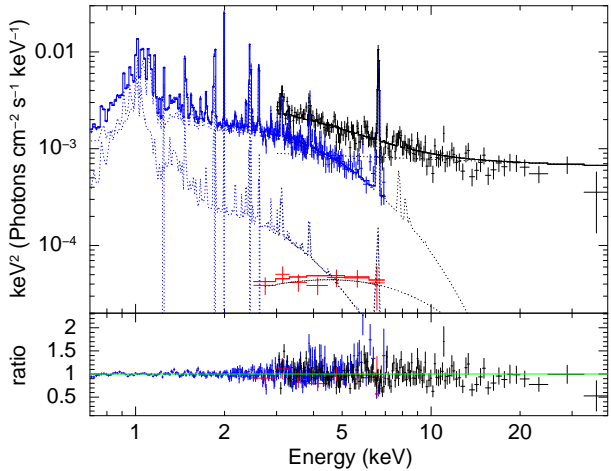}
\caption{
{\it NuSTAR} spectrum (black) within a circular region of 30\arcsec\ 
centered on the X-ray peak joint-fitted with the {\it Chandra} ICM (blue) and 
LMXB (red) spectra.  The dashed lines are the individual components 
discussed in the text.  The vertical dashed blue lines below 3\,keV 
are the Gaussian 
models, to take into account the line emission.  Beyond $\sim$5\,keV, 
the X-ray emission is dominated by the AGN-related activities (core, 
knots, and jet) and is completely dominated by these activities 
beyond $\sim$15\,keV.  
Error bars are at 1$\sigma$.
}
\label{fig:spec1}
\end{figure}

\begin{figure}
\includegraphics[width=2.5truein, angle=270]{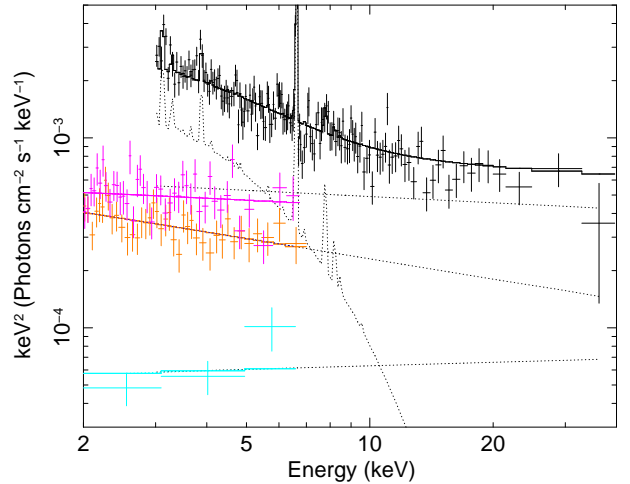}
\caption{
Same as Figure~\ref{fig:spec1} but the extra 
core (magenta; upper), jet (orange; middle), and HST-1 (cyan; lower) 
components are
also joint-fitted.  The three power-law dashed lines are models for these extra 
components best-fitted to the {\it NuSTAR} data.
The curved dashed line is the hotter ICM component.  For 
clarity, the cooler ICM and LMXB components and the residual panel are 
not shown.
}
\label{fig:spec2}
\end{figure}

\section{Overall Hard X-ray Emission from AGN Activities}

To constrain the overall hard X-ray emission from the AGN activities, 
which includes the unresolved X-ray core, knots, and the jet, we fitted 
the {\it NuSTAR} spectrum jointly with the {\it Chandra} ICM and LMXB 
spectra.  
All the spectra were grouped with a minimum of one
count per bin and were fitted using the $C$-statistic in the
the X-ray Spectral Fitting Package
({\tt XSPEC}).\footnote{http://heasarc.nasa.gov/xanadu/xspec/}
Errors of spectral parameters were determined by assuming $\Delta C 
= 2.706$ (90\% confidence) for one parameter of interest.
The {\it Chandra} ICM spectrum in the 
0.7--7.0\,keV range was well fitted with an absorbed two-temperature 
thermal model \citep[{\tt PHABS*(APEC+APEC)}; see also][]{RFM+15}.  We 
fixed the absorption at the Galactic value of $N_H = 1.94\times 
10^{20}$~cm$^{-2}$ \citep{KBH+05}, metallicity at solar value 
\citep{RFM+15}, and redshift at $z=0.004283$ \citep{Cap+11}.  The two 
temperatures and the two normalizations were free parameters.  We have 
also taken into account some residual ICM line emissions below 3\,keV by 
adding five {\tt GAUSSIAN} models with zero line width, although 
ignoring these will not have any impact on our results.

The {\it Chandra} LMXB spectrum between $\sim$2 and 7\,keV was fitted 
using a template given by \citet{RSK+14}, which is based on the 
broadband 3--100\,keV spectrum of M31 dominated by LMXBs and is 
consistent with the Galactic bulge emission 
\citep{KRC+07}.\footnote{Note that the first power-law index in the 
equation given by \citet{RSK+14} should be $-0.5$.} We fixed the shape 
of the adopted spectrum but varied its overall normalization.  Since the 
LMXB component in the {\it NuSTAR} spectrum is subdominant (see 
Figure~\ref{fig:spec1} below), increasing the normalization by a factor 
of two or using different LMXB models (e.g., a single power-law model or 
even a broken power-law template used by \citet{Wik+14b} to actually fit 
the harder high-mass binary dominated sources) gives essentially the 
same results.

The 3--40\,keV {\it NuSTAR} spectrum was joint-fitted with the {\it 
Chandra} ICM and LMXB spectra by including these two component models as 
well as a single power-law model to represent the combined emission from 
the core, knots, and the jet.  The same Galactic absorption model was 
also used for all the components.  We tied all the parameters for the 
ICM and LMXB models, except for the ICM component.  The ratio of the two 
{\tt APEC} normalizations is tied to that of the {\it Chandra} model 
but not the overall normalizations to allow for the cross-calibration 
uncertainty of {\it NuSTAR} and {\it Chandra} and the difference in 
response due to the much larger PSF of {\it NuSTAR}.  The difference in 
best-fit normalizations turns out to be small, with less than a 1\% 
difference when constraining the overall hard X-ray emission from the 
AGN activities (this section) and about 6\% when decomposing the core 
emission from the knots/jet (Section~\ref{sec:spec2}).

The best-fit spectra with all the model components are shown in 
Figure~\ref{fig:spec1}.  The best-fit power-law index of the overall 
AGN-related emission is $\Gamma = 2.12^{+0.12}_{-0.13}$ and the 
associated 20--40\,keV flux, which completely dominates the spectrum, is 
$7.7^{+1.1}_{-1.0} \times 10^{-13}$\,erg\,cm$^{-2}$\,s$^{-1}$.  The 
spectrum is well characterized by a single power law with no evidence of 
cut-off in energy below $\sim$100\,keV (3$\sigma$).\footnote{The lower 
limit of the cut-off energy is determined by the fitting of a cut-off 
power-law model (\tt cutoffpl).}

The 20--40\,keV flux measured is an order of magnitude lower than that 
determined by \citet{dBS+15} with the {\it Suzaku} observation taken in 
2006.  It is believed that the X-ray emission above 20\,keV detected 
near the end of 2006 was due to the HST-1 flare \citep{dBS+15, PFM+16}.  
The {\it Chandra} observations in 2017 show that the soft emission 
($\lesssim$10\,keV) from HST-1 has become fainter by a factor of 
$\sim$40 since 2006 November.  By simple scaling, the contribution from 
HST-1 in 20--40\,keV should be a factor of a few lower than that from 
the core.  This is indeed the case where HST-1 is a factor of six 
fainter than the core in 2017 measured in this band 
(Section~\ref{sec:spec2}).  Therefore, for the first time, we have 
detected hard X-ray emission above $\sim$10\,keV dominated by the core, 
which also gives the upper limit of its quiescent emission in hard 
X-rays.

\begin{deluxetable}{cccc}
\tabletypesize{\scriptsize}
\tablewidth{8.5cm}
\tablecolumns{5}
\tablecaption{Best-fit Spectral Results
\label{table:fit}
}
\tablehead{
\colhead{Name} &
\colhead{$\Gamma^{\rm a}$} &
\colhead{$K^{\rm b}$} &
\colhead{${f_{\rm 20-40\,keV}}^{\rm c}$}
}
\startdata
Core+HST-1+Jet & $2.12^{+0.12}_{-0.13}$  & $102^{+38}_{-29}$ & $7.7^{+1.1}_{-1.0}$\\
Core & $2.11^{+0.15}_{-0.11}$ & $63^{+28}_{-15}$ & $4.8^{+0.9}_{-1.0}$\\
HST-1 & [1.94] & $5.5^{+1.0}_{-0.9}$ & $0.7^{+0.1}_{-0.1}$\\
Jet & $2.36^{+0.16}_{-0.17}$ & $52^{+10}_{-10}$ & $1.8^{+0.9}_{-0.6}$
\enddata
\tablenotetext{}{\hspace{-3mm} {\bf Notes.}  The $C$-statistic in fitting 
the overall emission is $C = 1473$ for 1269 degrees of freedom, 
while $C = 2091$
when decomposing the different components for 1960 degrees of freedom.}
\tablenotetext{a}{Power-law photon index.  For HST-1, the parameter cannot be
constrained during the joint fitting and is fixed to the value determined
with {\it Chandra} alone.}
\tablenotetext{b}{Normalization in units of 
$10^{-5}$\,photons\,keV$^{-1}$\,cm$^{-2}$\,s$^{-1}$ at 1\,keV.}
\tablenotetext{c}{Flux in units of $10^{-13}$\,erg\,cm$^{-2}$\,s$^{-1}$.}
\end{deluxetable}

\section{Decomposing the Hard X-ray Emission of the Core, HST-1, and Jet}
\label{sec:spec2}

We can further constrain the individual components of the core, HST-1, 
and the rest of the knots and jet by joint fitting the corresponding 
{\it Chandra} spectra.  We model each of these components as a single 
power law.  The power-law indices of these components are tied for the 
{\it NuSTAR} and {\it Chandra} spectra.  HST-1 and jet emission is 
subdominant, and therefore their normalizations are tied for the {\it 
NuSTAR} and {\it Chandra} spectra.  The normalizations of the core 
component for {\it NuSTAR} and {\it Chandra} are not tied, to allow for 
calibration uncertainty.

The best-fit spectra with all the model components are shown in 
Figure~\ref{fig:spec2}.  The best-fit power-law index of the core is 
$\Gamma = 2.11^{+0.15}_{-0.11}$ and the associated 20--40\,keV flux 
measured with {\it NuSTAR} is $4.8^{+0.9}_{-1.0} \times 
10^{-13}$\,erg\,cm$^{-2}$\,s$^{-1}$, which is about 60\% of the overall 
AGN emission.  The normalization measured with {\it Chandra} is about 
10\% lower and smaller than the statistical uncertainty of about 20\%.  
Such consistency suggests that the assumed power-law models for 
different components are reasonable. The 20--40\,keV flux of HST-1 and 
the jet are about 16\% and 37\% of the core, respectively.  The best-fit 
parameters for the three components and the overall AGN emission are 
listed in Table~\ref{table:fit}.

\section{Discussion}
\label{sec:discussion}

We have detected hard X-ray emission from the M87 core, which can be 
readily compared with model predictions.  We focus on a few recent 
models with hard X-ray predictions based on the full SED constrained 
within $<$1\arcsec\ from the core \citep{PFM+16}.

\begin{figure}
\includegraphics[bb = 15 0 576 432, width=3.5truein, angle=0]{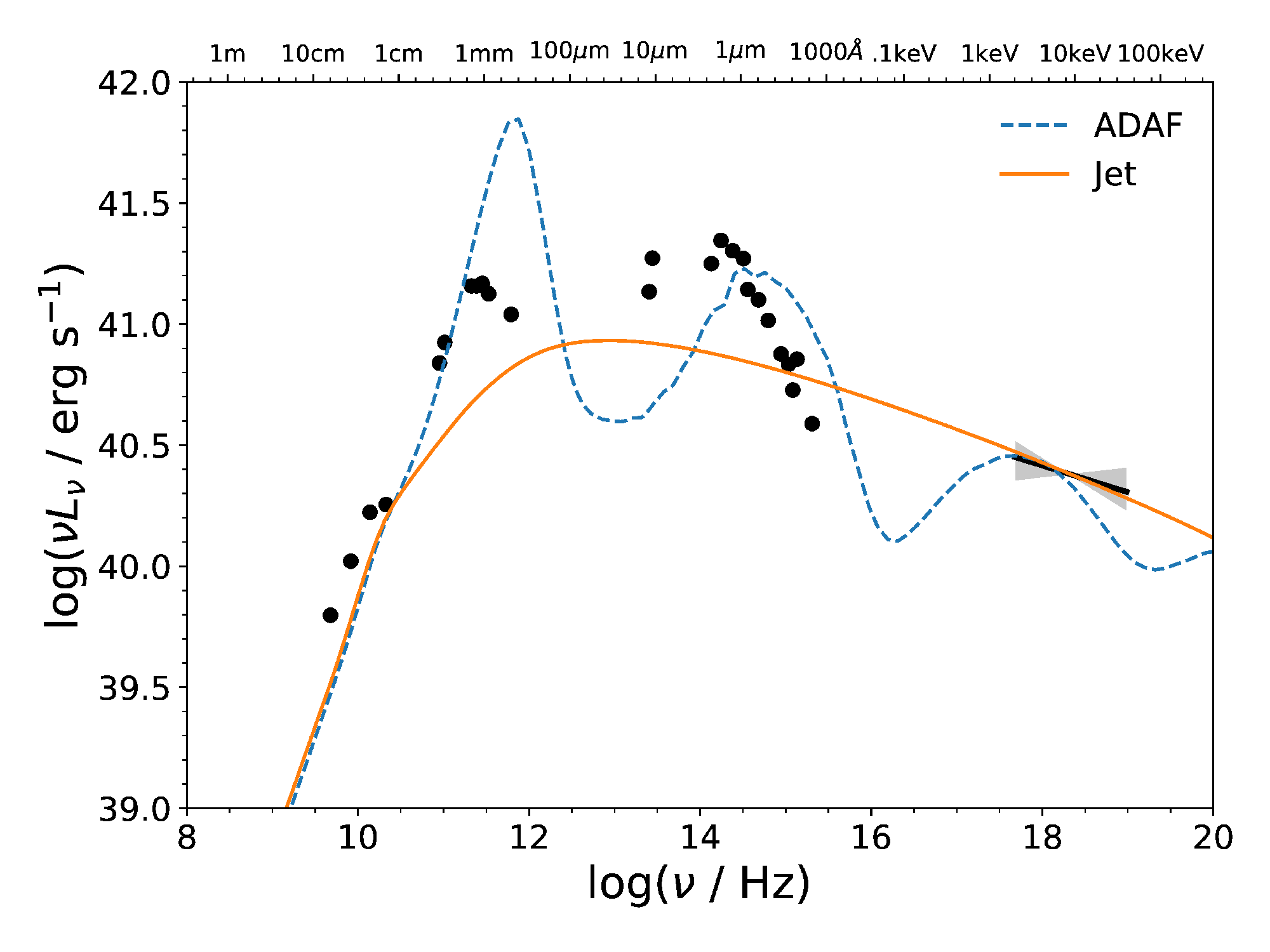}
\caption{
SED of the core with X-ray data taken from this work 
(90\% confidence region in photon index shown in gray)
and the rest taken 
from the 0\farcs4 quiescent data points of \citet{PFM+16}.  
The ADAF and jet models were taken from \citet{NSE14} and
renormalized to match the X-ray data.
}
\label{fig:models}
\end{figure}

Figure~\ref{fig:models} shows the SED of the M87 core, together with the 
ADAF-dominated and jet-dominated models of 
\citet{NSE14}. Both models were renormalized to match the X-ray data.  
In the past, it was difficult to distinguish these two models since both 
of them fit the SED well and predict similar slopes in soft X-rays below 
10\,keV \citep[see, e.g.,][]{NSE14}.  With {\it NuSTAR} data extending 
to 40\,keV, it is quite clear that the X-ray slope predicted by the 
ADAF-dominated model is too steep, while the X-ray slope and also the 
global SED are more consistent with the jet-dominated model.  Note that 
the featureless emission of a jet probably cannot explain the entire 
SED, particularly the IR to UV bump seen in the data.  A global fitting 
by varying the parameters of both the ADAF and jet components instead of 
simply renormalizing the models is needed to address the problem 
correctly.

Recently, with Faraday rotation measure observations, it has become 
clearer that the X-ray emission is more jet-dominated 
\citep{FWL16,LYX16}.  Our joint {\it NuSTAR} and {\it Chandra} 
observations provide additional insights into these models.  For example, 
our best estimated SED at 40\,keV ($\sim$$10^{19}$\,Hz) is $\nu F_{\nu} 
= 6.6^{+1.5}_{-1.3} \times 10^{-13}$\,erg\,cm$^{-2}$\,s$^{-1}$, which is 
consistent with the jet model predicted by \citet{PFM+16}.  However, by 
inspecting the X-ray slope in their Figure~3, the predicted photon index 
is $\Gamma \lesssim 2$, which is marginally flatter than our measured 
index of $2.11^{+0.15}_{-0.11}$. Comparing with model predictions by 
\citet{LYX16}, our result is marginally consistent with the 
jet-dominated model ($\approx$$5\times 
10^{-13}$\,erg\,cm$^{-2}$\,s$^{-1}$) and slightly higher than the 
ADAF-dominated model ($\approx$$3\times 
10^{-13}$\,erg\,cm$^{-2}$\,s$^{-1}$).  The measured photon index is also 
slightly more consistent with the jet-dominated model ($\approx$2.1) 
than the ADAF-dominated model ($\approx$2.2).  Note that the reason for 
not being able to completely rule out one model in favor of another 
(jet-dominated versus ADAF-dominated) is related to the theoretical 
uncertainties and freedoms allowed in the modeling.

In order to explain $\gamma$-ray and TeV emission from M87, the SED has 
been modeled using the synchrotron self-Compton (SSC) model that fits 
the softer X-ray ($\lesssim$10\,keV) and above $\gamma$-ray emission 
fairly well during the quiescent state of the core (Figure~3 in 
\citealt{dBS+15}, also, Figure~4 in \citealt{Abd+09}).  However, they 
overpredict the hard X-ray ($\sim$40\,keV) emission measured with {\it 
NuSTAR} by a factor of about three.  Their predicted power-law index 
($\sim$1.6) near 40\,keV is also significantly flatter than the measured 
value. {\it NuSTAR} is probing the transition from synchrotron dominant 
to (self) inverse Compton dominant emission around 10\,keV and is 
providing the key information on these VHE processes. Currently, the 
uncertainties are limited by the statistics of our data. A deep {\it 
NuSTAR} observation extending detection beyond 40\,keV, and simultaneous 
high angular resolution multi-wavelength observations, will provide 
further constraints on accretion models and VHE emission mechanisms.

\acknowledgments
We thank the referee for helpful comments.
This work was supported by NASA {\it Chandra} 
grant GO7-18085X and NASA {\it NuSTAR} grant NNX17GF12P.

\end{document}